\documentclass{revtex4}
\begin{document}

\title{A new operating mode in experiments searching for free neutron-antineutron oscillations based on coherent neutron and antineutron mirror reflections }
\author{V.V\,Nesvizhevsky$^1$, V.\,Gudkov\,$^2$, K.V.\,Protasov$^3$, W.M.\,Snow$^4$ and A.Yu.\,Voronin$^5$ \\
\medskip
{$^1$\,Institut Max von Laue -- Paul Langevin, 71 avenue des Martyrs, Grenoble, France-38042}\\
{$^2$\,Department of Physics and Astronomy, University of South California, South California, USA-29208}\\
{$^3$\,Laboratoire de Physique Subatomique et de Cosmologie, UGA-CNRS/IN2P3, Grenoble, France-38026}\\
{$^4$\,Department of Physics, Indiana University, 727 E. Third St., Bloomingston, Indiana, USA-47405}\\
{$^5$\,P.N. Lebedev Physical Institute, 53 Leninsky prospect, Moscow, Russia-119991}
}
%\maketitle
\medskip

\begin{abstract}
An observation of neutron-antineutron oscillations ($ n-\bar{n}$), which violate both $B$ and $B-L$ conservation, would constitute a scientific discovery of fundamental importance to physics and cosmology. A stringent upper bound on its transition rate would make an important contribution to our understanding of the baryon asymmetry of the universe by eliminating the post-sphaleron baryogenesis scenario in the light quark sector. We show that one can design an experiment using slow neutrons that in principle can reach the required sensitivity of $\tau_{n-\bar{n}}\sim 10^{10}s$ in the oscillation time, an improvement of $\sim10^4$ in the oscillation probability relative to the existing limit for free neutrons. This can be achieved by allowing both the neutron and antineutron components of the developing superposition state to coherently reflect from mirrors. We present a quantitative analysis of this scenario and show that, for sufficiently small transverse momenta of $n/\bar{n}$ and for certain choices of nuclei for the $n/\bar{n}$ guide material, the relative phase shift of the $n$ and $\bar{n}$ components upon reflection and the $\bar{n}$ annihilation rate can be small.
\end{abstract}
\maketitle
\medskip

The possible existence of neutron-antineutron ($n-\bar{n}$) oscillations is of fundamental interest for particle physics and cosmology. $n-\bar{n}$ oscillations would violate baryon number ($\Delta B=2$) and possesses many other implications for new physics \cite{Stueckelberg38,Sakharov67,Kuzmin70,Georgi74,Zeldovich76,Hooft76,MohapatraPRL80,Kazarnovskii80,Kuo80,Chang80,
MohapatraPLB80,Chetyrkin81,Dolgov81,Cowsik81,Rao82,Misra83,Rao84,Kuzmin85,Fukugita86,Shaposhnikov87,Dolgov92,Huber01,
Babu01,Nussinov02,Babu06,Dutta06,Berezhiani06,Bambi07,Babu09,Mohapatra09,Dolgov10,Gu11,Morrissey12,Arnold13,Canetti13,
Babu13,Gouvea14,Berezhiani16,Berezhiani18,Grojean18}.
Sensitive searches for $\Delta B=2$  processes, especially with $\Delta (B-L)=2$, such as $n-\bar{n}$, have started to attract more scientific attention. Cosmological arguments which use the well-known Sakharov criteria \cite{Sakharov67} to generate the baryon asymmetry of the universe starting from a $B=0$ condition require $B$ violation. A baryon asymmetry generated above the electroweak scale and conserving $B-L$, such as $\Delta B=1$ and $\Delta L=1$ proton decay $p\rightarrow\pi_0 e^+$, could be erased at the electroweak phase transition by sphalerons. Several theoretical models posses $\Delta B=2$ processes leading to $n-\bar{n}$ without giving $p$ decay \cite{Babu01,Nussinov02,Babu06,Dutta06,Babu09,Mohapatra09,Babu13,Arnold13,Berezhiani16}. An especially interesting class of models collectively referred as post-sphaleron baryogenesis (PSB) \cite{Babu13} can generate the baryon asymmetry below the electroweak scale. The type of experiment proposed in this paper can rule out PSB models operating in the light quark sector in combination with constraints on other consequences of the model from the LHC. Then one would be able to conclude that the generation of the baryon asymmetry within the Sakharov paradigm requires an understanding of electroweak sphaleron physics. New analysis of existing data to constrain $\Delta B=2$ processes have appeared from SuperK \cite{Gustafson15} and SNO \cite{Aharmim17}. The possibility of more sensitive $B$ violation searches in future underground detectors such as HyperK \cite{HyperK18} and Dune \cite{Dune14} is leading to new work on $\bar{n}A$ dynamics \cite{Golubeva18}. An observation of $n-\bar{n}$ would put stringent limits on CPT violation in the nucleon sector \cite{Babu15} within the low energy effective field theory (EFT) for CPT/Lorentz violation known as the Standard Model Extension \cite{Kostelesky11}. $n-\bar{n}$ would constrain long-range gauge fields coupled to $B-L$, where it would improve on present constraints from tests of the equivalence principle over a broad range of couplings and ranges \cite{Mohapatra16,Kamyshkov17}. Recent theoretical studies \cite{Berezhiani15,Gardner15,Gardner16,Fujikawa16,McKeen16} have clarified the subtleties involved in properly understanding the discrete symmetry transformations of a composite strongly interacting bound system like the $n$ and make it clear that the observation of $n-\bar{n}$ would indicate that the neutron is a Majorana particle \cite{BabuMajorana15}. A new $\Delta B=2$ process, $n-\bar{n}$ conversion, has been identified and described \cite{Gardner18}, and the possibility of $n$ -- mirror $n$ oscillations \cite{BerezhianiIJMP04,BerezhianiPRL06}, indicated as a possibility in experiments \cite{Serebrov08,Serebrov09,Altarev09,BerezhianiEPJC12} using Ultra Cold Neutrons (UCNs) \cite{Luschikov69,Steyerl69} is the subject of active studies \cite{BerezhianiPRD17}. Recent studies have investigated in greater depth the limits of the so-called quasifree condition for the evolution of the $n/\bar{n}$ amplitudes in external magnetic fields \cite{Young17}. These developments in theory and experiment show that the approach described in this paper for improving the sensitivity for $n-\bar{n}$ oscillations is of general interest to the physics community.

Experimental searches for $\Delta B=2$ processes involving $n$ have so far been conducted in two ways. One channel searches for $n-\bar{n}$ oscillations of free $n$ in an experimental setup where one tries to avoid all Standard Model interactions of the $n/\bar{n}$ with matter and external magnetic fields which shift the relative energy between $n$ and $\bar{n}$ by $\Delta E$ and can suppress the oscillation rate. Despite the fact that $\Delta E \gg \varepsilon$ ($\varepsilon$ is the off-diagonal mixing term in the effective Hamiltonian for the $n/\bar{n}$ two-state system), the oscillation rate is not greatly suppressed if the observation time $t$ is short compared to $\Delta E/\hbar$ ($\hbar$ is the reduced Planck constant). In this so-called quasifree regime, the relative phase shift between the $n$ and $\bar{n}$ states from the time development, $e^{-\Delta Et/\hbar}$, is small enough that the oscillation probability still grows quadratically with $t$ for short enough observation times. The other channel searches for $n-\bar{n}$ oscillations of the neutrons bound in nuclei, where the rate is suppressed by a very large $\Delta E$ \cite{Kuo80,Dover83,Dover89,Gal00,Chung02,Friedman08,Abe15}. Still, the very large number of neutrons, which can be observed in large volume of low-background underground detectors makes this the most sensitive search mode at present. However, the interpretation of the results depends somewhat on models of $\bar{n}$ annihilation in nuclei and on models of the branching ratios for the many different reaction products from the $\bar{n}$ annihilation. Also, this process is not physically equivalent to free $n$ oscillations since there are additional $\Delta B=2$ processes which can happen inside a nucleus but not for a free $n$. Therefore, the bounds from these two processes should be treated as complementary.

We propose and analyze a new version of $n-\bar{n}$ experiment: an (almost) free neutron oscillation search in which we allow slow $n/\bar{n}$ to reflect from $n/\bar{n}$ optical mirrors. Although reflection of UCNs from trap walls was considered already in 1980 \cite{Kazarnovskii80,Chetyrkin81,Yoshiki89,Yoshiki92} for experiments constraining $\tau_{n-\bar{n}}$, we extend this approach to much higher energy, point out conditions for suppressing the phase difference for $n$ and $\bar{n}$ at reflection and underline the importance of setting low transverse momenta of $n/\bar{n}$ and making certain choices for the nuclei composing the guide material. We show that, over a broad fraction of phase space acceptance of a $n/\bar{n}$ guide, the probability of coherent reflection of $n/\bar{n}$ from the guide walls can be quite high, the relative phase shift between $n$ and $\bar{n}$ can be quite small, and the theoretical uncertainties in the calculation of the experimental sensitivity are small. We show that such an experimental mode can relax some of the constraints on free $n$ oscillation searches and in principle allows us to achieve a much higher sensitivity compared to approaches in which reflections of the $\bar{n}$ component of the amplitude are not allowed. This approach preserves both the very low backgrounds that have been achieved in free $n$ oscillation searches and the ability to confirm a nonzero signal by the application of a small external magnetic field which further splits the $n$ and $\bar{n}$ states by $\Delta E=2\mu B$. However, it does not require the same level of detail in the understanding of the $\bar{n}$ dynamics and the subsequent annihilation products needed to interpret the underground detector $\bar{n}$ annihilation experiments.

A general expression for the $n-\bar{n}$ oscillation probability \cite{Kerbikov03} resembles the well known equation for the neutral kaon oscillations. For any practical observation time, $e^{-\Gamma_{\beta}t}  \approx 1$ ($\Gamma_{\beta}$ is the $n$ $\beta$-decay width) and $\omega t\ll 1$ ($\omega$ is the oscillation frequency). Thus, it reduces to $P_{n\to \bar{n}}(t)\approx {\varepsilon}^2 e^{-\frac{\Gamma_{\alpha}t}{2}}t^2$, where $\Gamma_{\alpha}$ is the $\bar{n}$ annihilation width, $\varepsilon = E_n-E_{\bar{n}}$, $E_n$ and $E_{\bar{n}}$ are the energies of the $n$ and $\bar{n}$. Therefore, for the optimum observation time $t_0=4/\Gamma_{\alpha}$, the corresponding maximum $n-\bar{n}$ oscillation probability is
\begin{equation}\label{probability of oscillations}
P_{n \to \bar{n}} \approx 2.1 (\frac{\varepsilon}{\Gamma_{\alpha}})^2.
\end{equation}
Neglecting annihilation, it would reduce to the quasifree limit expression $P_{n \to \bar{n}} \approx (t / \tau_{n \to \bar{n}})^2$, where $\tau_{n \to \bar{n}} = 1/{| \varepsilon |}$ (for natural units $\hbar = c = 1$) is the oscillation time. As a few annihilation events suffice for a positive signal for the very low backgrounds that can be reached, the experimental figure of merit is $F\approx N t^2$ with the total number $N$ of $n$.

The best constraint on $\tau_{n\to\bar{n}}$ using free $n$ used an intense cold neutron beam at the Institute Laue-Langevin (ILL) \cite{Baldo94} with some techniques developed in earlier experiments \cite{Fidecaro85,Bressi90}. An ambitious project at a projected fundamental physics beamline at the European Spallation Source (ESS) \cite{Phillips15} proposes an analogous scheme with advanced parameters to increase the sensitivity by a factor of $G\approx 10^2-10^3$. It would require a large solid angle of neutron extraction from the ESS source. We argue that the sensitivity can be improved if one allows the reflection of $n/\bar{n}$ from $n/\bar{n}$ guide walls. This can increase the observation time $t$ and also improve the counting statistics. Some literature on the subject gives the incorrect impression that the coherence of the $n/\bar{n}$ amplitude is always destroyed upon contact with matter. This is not true for a coherent neutron reflection from a surface. As long as the $n$ is not "observed" (annihilated) and the phase difference between the $n$ and $\bar{n}$ components of the amplitude upon reflection is small enough, the sensitivity for the $\bar{n}$ component continues to grow with time just as it does in the quasifree limit. We give a formalism and estimations for an experiment at PF1B beam \cite{Abele06} at ILL as an example. Greater sensitivity could be achieved at other neutron sources/guides. A combined design including the extraction of neutrons through a large solid angle as projected in \cite{Phillips15} could lead to additional improvements. To estimate the sensitivity for different configuration and neutron sources, one can use standard neutron optical calculations with the formalism developed in this paper.

Soon after the discovery of the neutron \cite{Chadwick32}, Enrico Fermi introduced a point-like $n$-nuclear ($nA$) pseudo-potential \cite{Fermi36} for description of coherent scattering of slow neutrons: $U(\vec{r})=((2\pi{\hbar}^2)/m) b_{nA} \delta\vec(r)$, with $m$ the reduced neutron mass and $b_{nA}$ the complex scattering length. Then the interaction of $n$ with matter can be described using formal perturbation theory with complex optical (Fermi) potential $U(r)=((2\pi{\hbar}^2)/m) (\rho/\mu) b_{nA}$, with $\rho$ the mass density of material and $\mu$ the atomic mass. The potential $U(r)$ for composite materials is simply the weighted sum of potentials from the different nuclei (which were measured with the accuracy of $1\%$ or better). It is well known that a small grazing angle reflection of $n$ from materials with positive potential allows the construction of $n$ guides \cite{Maier63}. This is applied also to $\bar{n}$ reflections with only difference that $Im(U(r))$ values are always large due to $\bar{n}$ annihilation. The important parameters for the analysis of $n-\bar{n}$ oscillation experiments are the probabilities of $n$ and $\bar{n}$ reflection per bounce ($\rho_n$ and $\rho_{\bar{n}}$) and the relative phase shift between the $n$ and $\bar{n}$ wave functions per bounce $\Delta\varphi_{n\bar{n}} = \varphi_{n} - \varphi_{\bar{n}}$, where $\varphi_n$ and $\varphi_{\bar{n}}$ are phase shifts of the wave function inside the bulk for $n$ and $\bar{n}$. The parameters $\rho_n$, $\rho_{\bar{n}}$, $\varphi_n$, $\varphi_{\bar{n}}$ depend on the energy of $n/\bar{n}$ transverse motion ($e_n$ and $e_{\bar{n}}$) and on the optical potentials of the wall material $U_n=V_n+iW_n$ and $U_{\bar{n}}=V_{\bar{n}}+iW_{\bar{n}}$ for the $n$ and $\bar{n}$, where $V_n$ and $V_{\bar{n}}$ are real parts, and $W_n$ and $W_{\bar{n}}$ are the imaginary parts. The reflection probabilities $\rho_n$ and $\rho_{\bar{n}}$ are:
\begin{eqnarray}\label{probability of antineutron reflection}
\rho_n=1, \rho_{\bar{n}}=1-\frac{4kk_{\bar{n}}^{''}}{(k+k_{\bar{n}}^{''})^2+(k_{\bar{n}}^{'})^2}, k_{\bar{n}}^{'}-ik_{\bar{n}}^{''}=\sqrt{ 2m(V_{\bar{n}}-iW_{\bar{n}}-e)},
\\
k_{\bar{n}}^{'}=\sqrt{m(\sqrt{(V_{\bar{n}}-e)^2+(W_{\bar{n}})^2}+(V_{\bar{n}}-e))}, k_{\bar{n}}^{''}=\sqrt{m(\sqrt{(V_{\bar{n}}-e)^2+(W_{\bar{n}})^2}-(V_{\bar{n}}-e))},
\end{eqnarray}
with $k_{\bar{n}}$ the complex momentum of $\bar{n}$ inside the wall. For the case of $n-\bar{n}$ oscillation experiments, we are interested only in specular reflection since its probability can reach $\sim 99.9\%$ even for UCNs \cite{Nesvizhevsky06,Nesvizhevsky07}, and in $n/\bar{n}$ with transverse energies $e=(k^2)/(2m)$ small compared to $V_n$ and $V_{\bar{n}}$ ($e\ll V_n, e\ll V_{\bar{n}}$) but comparable to $W_{\bar{n}}$ ($e\sim W_{\bar{n}}$). Also, based on existing theoretical analysis of the $\bar{n}A$ data for various nuclei, we know that $W_n\ll V_n, W_{\bar{n}}\ll V_{\bar{n}}$ and $W_n\ll W_{\bar{n}}$. It is remarkable that the reflection probability $\rho_{\bar{n}}$ (\ref{probability of antineutron reflection}) for the cases of weak ($|V_{\bar{n}}|\gg |W_{\bar{n}}|$), intermediate ($|V_{\bar{n}}|\sim|W_{\bar{n}}|$), and strong ($|V_{\bar{n}}|\ll |W_{\bar{n}}|$) absorptions are close to unity and quite insensitive to the variation in the magnitude of $U_{\bar{n}}$. This is consistent with known facts for analogous physical systems where strong losses do not destroy quantum coherence, for example in the reflection of polarized light from a metal mirror. Using these conditions, we can simplify eq. (\ref{probability of antineutron reflection}) as follows: $k_{\bar{n}}^{'}\approx \sqrt{2mV_{\bar{n}}}$, $k_{\bar{n}}^{''}\approx \sqrt{m(W_{\bar{n}}^2/({2V_{\bar{n}}}))}$, and obtain, in a first order approximation, expressions for $1-\rho_{\bar{n}}$ and $\Delta\varphi_{n{\bar{n}}}$ as:
\begin{equation}
\label{reduced probability and phase shift of antineutron reflection}
1-\rho_{\bar{n}}\approx\frac{4kk_{\bar{n}}^{''}}{(k_{\bar{n}}^{'})^2},
\Delta\varphi_{n{\bar{n}}}\approx\frac{2k}{k_nk_{\bar{n}}^{'}}(k_n-k_{\bar{n}}^{'}), \Delta\varphi_n=\arctan({-\frac{2kk_n}{k^2-k_n^2}}), \Delta\varphi_{\bar{n}}=\arctan({-\frac{2kk_{\bar{n}}^{'}}{k^2-(k_{\bar{n}}^{'})^2-(k_{\bar{n}}^{''})^2}}).
\end{equation}

The low energy $\bar{n}A$ scattering can be described by a scattering length $b_{\bar{n}A}$. The similarity between $\bar{p}$ and $\bar{n}$ low-energy scattering on nuclei allowed the authors of \cite{Batty01} to suggest the fitting formula for $b_{\bar{n}A}$:
\begin{equation}\label{anti scattering length}
b_{\bar{p}(\bar{n})A}=(1.54A^{1/3}-i1.0)fm,
\end{equation}
where $A$ the atomic number. $Im(b_{\bar{p}(\bar{n})A})$ is estimated as $\sim 1fm$ from the diffusive tail of an effective $\bar{n}A$ potential which is rather similar for all nuclei. $Re(b_{\bar{p}(\bar{n})A})$ is proportional to the nuclei radius. The $\bar{n}A$ interaction is restricted to the nuclear surface and is therefore practically insensitive to the internal structure of the nucleus \cite{Protasov00,Batty01}. There is also no $\bar{n}$ counterpart to compound $nA$ resonances, which greatly complicate the first-principles calculation of $b_{nA}$. $U_{\bar{n}}$ is therefore much easier to calculate than $U_n$. The calculated values of scattering lengths, $U_{\bar{n}}$ potentials and lifetimes of $\bar{n}$ on a surface of corresponding materials are presented in Table \ref{Table1} \footnote{Note that the value of $\bar{n}Cu$ optical potential published in \cite{Kerbikov03,Kerbikov04} is inconsistent with the corresponding scattering length, published in the same paper, and is probably wrong}.
\begin{table}
 \centering
 \begin{tabular}{|c|l|l|l|}
  \hline

Element&$b_{\bar{n}A}$ [fm]& $U_{\bar{n}}$ [neV]& $\tau_{\bar{n}}$ [s]\\
  \hline
  C & $3.5-i$ & $103-i29$ & $1.7$ \\
  \hline
  Mg & $3.5-i$ & $39-i11$ & $1.0$ \\
  \hline
  Si & $3.7-i$ & $48-i13$ & $1.2$ \\
  \hline
  Ni & $4.7-i$ & $111-i24$ & $2.3$ \\
  \hline
  Cu & $4.7-i$ & $104-i22$ & $2.2$ \\
  \hline
  Zr &	$5.3-i$& $59-i11$ &	$1.8$ \\
  \hline
  Mo &	$5.3-i$& $89-i16$& $2.3$ \\
  \hline
  W&	$6.5-i$&	$106-i16$&	$3.0$ \\
  \hline
  Pb&	$6.7-i$&	$57-i8.6$&	$2.3$ \\
  \hline
  Bi&	$6.7-i$&	$49-i7$&	$2.1$ \\
  \hline
 \end{tabular}
 \caption{Parameters which characterize the interaction of $\bar{n}$ with different materials: $b_{\bar{n}A}$ (the scattering length), $U_{\bar{n}}$ (the complex optical potential for this material), $\tau_{\bar{n}}$  (the time of storage of $\bar{n}$ with close-to-zero vertical energy on a horizontal surface in the gravitational field of the Earth). Calculations for all elements are averaged over the natural isotopic compositions.
 %Note that the value of $\bar{n}Cu$ optical potential published in [36], [57] is inconsistent with the corresponding scattering length, published in %the same paper, and is probably wrong.
 } \label{Table1}
 \end{table}
  
It should be noted that eq. (\ref{anti scattering length}) is a good approximation for mean $b_{nA}$ values as noticed in \cite{Peshkin71} and used in \cite{NesvizhevskyPRD08} for constraining short-range forces. Moreover, the regular dependence of $b_{\bar{n}A}$ on nuclei gives the opportunity to choose the composition of materials with equal real parts of $U_n$ and $U_{\bar{n}}$ potentials. For example, the isotopic composition $^{184}W(87.7\%)+^{186}W(12.3\%)$ results in $U_n \sim U_{\bar{n}} \sim 106neV$.

Consider a ballistic $n$ guide \cite{Abele06} consisting of two following parts. Its cross sectional area $s=hd$ ($\sim 10^2 cm^2$) at the $1^{st}$ (upstream) section increases along its length; let $h$ be its height and $d$ its width at the entrance, $H$ its height and $D$ its width at the exit, and $l$ the length. In the $2^{nd}$ part, the cross sectional area $S=HD$ ($\sim 10^4 cm^2$) is constant over its length with sizes $H$ and $D$, and the length $L$. Since the neutrons, which strike the wall in the extending part see the wall recede in their frame, these collisions lower the transverse components of the neutron velocity. We assume $|v_{hor}|, |v_{vert}|< 2v_{crit}^{Ni}$ at the entrance to the guide, with $v_{crit}^{Ni}\approx 7m/s$. In accordance with Liouville's theorem, $|v_{hor}|\cdot |v_{vert}|<(2v_{crit})^2\frac{dh}{DH}$ at the exit of a properly designed adiabatic guide. A typical high value of $U_{\bar{n}}$ in Table \ref{Table1} is $>100 neV$ corresponding to critical velocities of $>4m/s$. To analyze the problem in the low-energy limit, we assume $\overline{{|v_{hor}|}} \sim \overline{{|v_{vert}|}} \sim 1 m/s$. These two conditions are met, if the guide cross sectional area is expanded by at least the factor of $\frac{DH}{dh}\sim 49$. For practical arrangements, one would use a few superimposed flat guides and design a guide shape that mixes horizontal and vertical velocities of the neutrons. To account for this option, we reduce the size of $D$ to $1 m$. Note that the diverging part contributes to the $n-\bar{n}$ sensitivity provided the $n/\bar{n}$ incidence angles are small. We select copper as a material for this analysis because $Re(b_{\bar{n} Cu})$ is large, $Im(b_{\bar{n} Cu})$ is relatively small, and because $Cu$ has been successfully used in the past for $n$ mirrors. Another good example is tungsten with an adjusted isotopic composition since it provides longer storage time of $\bar{n}$ on the surface.

Because of the importance of gravity, we consider the interactions of $n/\bar{n}$ with horizontal and vertical walls separately. The frequencies of $n/\bar{n}$ collisions with horizontal walls and bottom (provided $n/\bar{n}$ never touch the top) are $f_{hor}=|v_{hor}|/D$ and $f_{vert}=g/(2|v_{vert}|)$. The $\bar{n}$ lifetimes associated with side-walls and bottom are $\tau_{hor}^{\rho,\bar{n}}=1/(f_{hor}(1-\rho_{\bar{n}}))$ and $\tau_{vert}^{\rho,\bar{n}}=1/(f_{vert}(1-\rho_{\bar{n}}))$; the annihilation probability is given by eq. (\ref{probability of antineutron reflection}). A lifetime for the accumulation of a phase shift of $1 rad$ between $n$ and $\bar{n}$ amplitudes associated with side-wall and bottom collisions are $\tau_{hor}^{\Delta\varphi,\bar{n}}=1/(f_{hor}\Delta\varphi_{n\bar{n}})$ and $\tau_{vert}^{\Delta\varphi,\bar{n}}=1/(f_{vert}\Delta\varphi_{n\bar{n}})$; the phase shift is given by eq. (\ref{reduced probability and phase shift of antineutron reflection}). Thus the accumulation time for a phase difference $\Delta\varphi_{n\bar{n}}$ of $1 rad$ due to the side-wall collisions, $\tau_{hor}^{\Delta\varphi, \bar{n}}$,
is $32 s$ for $Cu$ and very long for isotopically adjusted $^{184+186}W$. The characteristic accumulation time for a phase difference $\Delta\varphi_{n\bar{n}}$ of $1 rad$ due to bottom collisions,
\begin{equation}\label{vertical phase}
\tau_{vert}^{\Delta\varphi , \bar{n}}=\frac{\overline{|v_{vert}|}}{g}\frac{\sqrt{V_nV_{\bar{n}}}}{\sqrt{\overline{e_{vert}}}\cdot |\sqrt{V_n}-\sqrt{V_{\bar{n}}} |},
\end{equation}
is $7.3 s$ for $Cu$ and much longer for $^{184+186}W$. For $Cu$, $\tau_{vert}^{\Delta\varphi , \bar{n}} \ll \tau_{hor}^{\Delta\varphi , \bar{n}}$ because of gravity. A proper mixture of materials/ isotopes for the $n/\bar{n}$ guide walls (as for $^{184+186}W$) would increase $\tau_{vert}^{\Delta\varphi , \bar{n}}$ due to the term $((\sqrt{V_n}-\sqrt{V_{\bar{n}}})\to 0)$ in the denominator of eq. (\ref{vertical phase}). The annihilation time due to the side-wall collisions, $\tau_{hor}^{\rho, \bar{n}}$,
is $11 s$ for $Cu$ and $15 s$ for $^{184+186}W$. All these time scales are large enough to neglect corresponding processes as sources of the possible incoherence. The annihilation time due the bottom collisions,
\begin{equation}\label{horizontal lifetime}
\tau_{vert}^{\rho, \bar{n}}=\frac{\overline{|v_{vert}|}}{g}\frac{(V_{\bar{n}})^{3/2}}{W_{\bar{n}}\sqrt{\overline{e_{vert}}}},
\end{equation}
is $2.2 s$ for $Cu$ and $3.1 s$ for $^{184+186}W$. Not surprisingly, annihilation of $\bar{n}$ in the accumulated reflections from the bottom is the limiting factor; because of the presence of gravity, such collisions cannot be avoided in a horizontal guide. The effect of gravity on $n-\bar{n}$ experiment became less important when we use a parabolic $n/\bar{n}$ guide. Then $g$ in eq. (\ref{horizontal lifetime}) is replaced by a smaller $g_{eff}$ value, and $\tau_{ver}^{\rho,\bar{n}}$ increases. For the maximum effect, the parabolically-curved guide could follow a typical trajectory of a $n/\bar{n}$ with a mean velocity in the beam. The above estimations neglect the energy corrections that is justified by the weak dependence of $\tau_{vert}^{\rho,\bar{n}}$ on $e_{vert}$ due to the approximate compensation of larger frequency of bounces of $n/\bar{n}$ by smaller probability of $\bar{n}$ annihilation per bounce. Therefore, the $n-\bar{n}$ experiment sensitivity is essentially defined by one value (eq. (\ref{horizontal lifetime})) specific for each $n-\bar{n}$ guide material. This simplification helps also to estimate the impact of uncertainties in $b_{\bar{n}A}$ values to the experiment sensitivity. As $\tau_{vert}^{\rho,\bar{n}} \sim 1/W_{\bar{n}}$ (eq. (\ref{horizontal lifetime})), a typical $10-20\%$ error in the estimation of $W_{\bar{n}}$ gives only a $10-20\%$ error in the calculation of $\tau_{vert}^{\rho,\bar{n}}$. If the observation time is shorter than the optimum value, the impact is even smaller. The impact of uncertainty in the $V_{\bar{n}}$ is negligible by comparison.

The above arguments treat $n/\bar{n}$ motion semiclassically. However, the results coincide with quantum expressions \cite{Voronin06,Voronin11} in the low-energy limit. The scattering length of a cold $n/\bar{n}$ is $a=1/k_{n,\bar{n}}=1/\sqrt{2m(V_{n,\bar{n}}+W_{n,\bar{n}})}$ \cite{Voronin11}. The effective horizontal momentum of a $n/\bar{n}$ in a box with a size $D$ is $k\approx \pi j/(D-2/k_{n,\bar{n}})$, where $j$ is quantum number of the box-like state. The horizonal energy levels shift is $\Delta E_{hor}\approx 4\varepsilon_{hor} a/D=4\varepsilon_{hor}/(Dk_{n,\bar{n}})$. The vertical energy levels shift is $\Delta E_{ver}=mga=mg/k_{n,\bar{n}}$ \cite{Voronin11}. This expression is energy-independent and consistent with the arguments given above. It is easy to show that quantum expressions for the timescales $\tau_{hor}^{\Delta\varphi,\bar{n}}=1/\omega_{hor}$, $\tau_{hor}^{\rho,\bar{n}}=1/\Gamma_{a,hor}$, $\tau_{ver}^{\Delta\varphi,\bar{n}}=1/\omega_{ver}$, $\tau_{ver}^{\rho,\bar{n}}=1/\Gamma_{a,ver}$ also coincide with the semiclassical expressions obtained above. The whole analysis in this paper was initiated by the observation that while $\tau_{vert}^{\rho,\bar{n}}$ is short for $e_{n/\bar{n}}$ close to $V_{n/\bar{n}}$ \cite{Kerbikov04}, it is longer for smaller energies. However, even the limit of gravitational quantum states \cite{Nesvizhevsky02} does not bring an additional improvement in the observation time as the annihilation time saturates.

The total number of neutrons at PF1 instrument was $3\cdot 10^{18}$ in the previous free $n$ experiment \cite{Baldo94}. The number of $n$ per year at PF1B instrument could be 4-5 times larger. A gain factor due to the increased path could be $\sim 10^2$ for $t\sim 1 s$, and $\sim 10^4$ for $t\sim 10 s$. Any project has to optimize the sensitivity relative to the $n/\bar{n}$ beam geometry, neutron spectrum, the budget and spatial constrains; such considerations are beyond the scope of this paper. Assuming that detection of a couple of annihilation events in a background-free experiment means the observation of $n/\bar{n}$ oscillations, one can estimate the overall sensitivity with the following simple formula:
\begin{equation}\label{overall accuracy}
\tau_{n \to \bar{n}}\sim\frac{\sqrt{FT}}{\Gamma_a},
\end{equation}
where $F$ the total $n$ flux, and $T$ the experiment duration. For the same $F$, one prefers a softer $n$ spectrum to decrease the experiment length, thus taking full advantage of this operation mode.

The gravitational constraints from wall reflections are obviously no longer relevant for a vertical $n/\bar{n}$ guide. Consider an upwards-directed fountain of very cold neutrons (VCNs). In this case, the problem is turned into a great advantage: the observation time is long and can include both the rise and the fall. Basing on the estimations given above, annihilation and phase shifts in the guide walls are low and can be neglected, thus providing practically pure quasifree limit conditions. To increase VCN fluxes, one could design a dedicated large-surface VCN source based on fluorinated nano-diamond reflectors \cite{NesvizhevskyCarbon18,NesvizhevskyPRA18}. For a typical VCN velocity of $\sim 50 m/s$, the fountain height is $\sim 125 m$ and the observation time is $\sim 10 s$. For an estimated flux density of $\sim 10^7-10^8 VCN/cm^2/s$, we approach the sensitivity estimated above, however largely gain in the experiment size and decrease the cost.

Finally, we emphasize that recent and future progress in our understanding of the $\bar{n}A$ optical potential and particularly in quantifying the uncertainties of real and imaginary parts of the optical potential could significantly improve the sensitivity of $n-\bar{n}$ experiment and lower its cost through better choice of material for the $n/\bar{n}$ guide also through more precise optimization of this expensive experiment. Both theoretical and experimental efforts in the understanding of $\bar{n}$ annihilation by nuclei are highly encouraged.

This publication is funded by the Gordon and Betty Moore Foundation to support the work of V.V.N. and W.M.S. The authors are grateful to the participants of INT-17-69W Workshop "Neutron-antineutron oscillations: appearance, disappearance and baryogenesis" in Seattle, USA, held on 23-27 October 2017, as well as to our colleagues from GRANIT collaboration. V.G. is grateful for support to the U.S. Department of Energy, Office of Science, Office of Nuclear Physics program under Award Number DE-SC0015882. W.M.S. acknowledges support from NSF PHY-1614545 and from the Indiana University Center for Spacetime and Symmetries.

%For tables use syntax in table~\ref{tab-1}.
%\begin{table}
%\centering
%\label{tab-1}       % Give a unique label
%% For LaTeX tables you can use
%\begin{tabular}{lll}
%\hline
%first & second & third  \\\hline
%number & number & number \\\hline
%\end{tabular}
%% Or use
%\vspace*{5cm}  % with the correct table height
%\end{table}
%
% BibTeX or Biber users please use (the style is already called in the class, ensure that the "woc.bst" style is in your local directory)
% \bibliography{name or your bibliography database}

\end{document}